\documentstyle[floats,psfig,prc,aps,manuscript]{revtex}

\begin{document}

\title{ 
Strongly damped nuclear collisions: zero or first sound ? 
      } 

\author{ 
A.B. Larionov$^{1,2}$, J. Piperova$^3$, M. Colonna$^4$, M. Di Toro$^4$ 
       }

\address{ $^1$ Institut f\"ur Theoretische Physik, 
          Universit\"at Giessen,D-35392 Giessen, Germany }

\address{ $^2$ Russian Research Center "I.V. Kurchatov 
          Institute",123182 Moscow, Russia }

\address{ $^3$ Institute for Nuclear Research and Nuclear Energy, 
          Sofia, Bulgaria }  

\address{ $^4$ Laboratori Nazionali del Sud, Via S. Sofia 44,
          I-95123 Catania, Italy\\ and University of Catania }

\maketitle

\begin{abstract}

The relaxation of the collective quadrupole motion in the 
initial 
stage of a central heavy ion collision at beam energies 
$E_{lab}=5\div20$ AMeV is studied within a microscopic kinetic
transport model.
The damping rate is shown to be a non-monotonic function of $E_{lab}$ 
for
a given pair of colliding nuclei. This fact is 
interpreted as a 
manifestation of the zero-to-first sound transition in a 
finite nuclear 
system.

\end{abstract}

\vspace{0.5cm}

\section{ Introduction }

The phenomenon of strong dissipation in the collective 
motions of 
heated nuclear systems is a challenging problem for the 
nuclear transport
theory. Following the analogy with other Fermi liquids,
like the liquid $^3$He 
\cite{AK,abe66}, one
expects also in nuclear systems two types of collective 
modes.

At small temperatures the mean field dominated modes 
should exist, from the possible violatation of
the local thermal equilibrium due to the strong Pauli 
blocking which inhibits two-body collisions
(the Landau zero sounds). This is actually the main nature of
isoscalar giant resonances, where indeed the collective energy
is essentially given by the amount needed to deform the
Fermi sphere in momentum space \cite{ber78,rs80}.
Isovector giant resonances are also mean field modes, this
time of plasmon type since we can have both a distortion and a 
shift of the neutron (proton) Fermi spheres.

At high temperatures the usual hydrodynamical collective 
modes (first sounds)
should propagate \cite{KPS96}. In an infinite Fermi 
liquid with strong
repulsion (Landau parameter $F_0 \gg 1$) the two sounds 
are clearly 
distinguishable, since the damping rate has a maximum as 
a function of the
temperature \cite{AK,abe66}. However, in nuclear matter 
the transition
between the two sounds is expected to be smeared-out 
since the Landau 
parameter $F_0$ is small in absolute value \cite{LDK}.

The presence of this transition is still an open problem in 
nuclear dynamics. Only the isovector giant dipole resonances
can be experimentally studied with sufficient accuracy in 
heated nuclei. The point is that for this kind of two-component
mode the transition has some special features that make it 
difficult to observe a clear signature \cite{LCBD99}.

Aim of this work is to study the transition in the collective
nuclear dynamics looking at
the evolution of the damping mechanism of large
amplitude quadrupole oscillations in fusion processes,
in a microscopic kinetic model.
Similar attempts have been recently performed in fission dynamics 
studies \cite{Wilc96,Don99}. We will also follow the temperature
dependence of the attenuation of the collective mode.

In the ref.s \cite{Wilc96,Don99} a reduced 
friction coefficient
\begin{equation}
                  \beta = {1 \over E_{kin}^{coll}} 
                          \left( d E \over d t 
\right)_{diss}
                                          \label{beta},
\end{equation}
where $E_{kin}^{coll}$ is the collective kinetic energy of the 
dinuclear system (DNS)
and $(d E / d t)_{diss}$ is the dissipation rate of the 
total collective
energy $E = E_{kin}^{coll} + E_{pot}$, has been extracted from 
the measurements 
of the prescission neutron multiplicity in fast fission 
reactions.
It was shown that in the 
temperature region
$T = 2\div3$ MeV the reduced friction coefficient $\beta$ 
is very large
($\beta \sim 10\div100 \cdot 10^{21}$ s$^{-1}$), result
not explained
with a one-body dissipation mechanism only.

In the present work we have performed some kinetic transport 
studies of the nuclear dynamics. Transport equations describe
a self-consistent mean field dynamics coupled to two-body
collisions and so we expect to see in a natural way the transition
between the two sound propagations. 
We use
the 
Boltzmann-Nordheim-Vlasov
(BNV) \cite{BNV} procedure to simulate the phase space dynamics,
which has been quite successful in predicting mean properties
of heavy ion collisions at medium energies. In particular
we study central nucleus-nucleus collisions leading to fusion
at beam
energies $E_{lab} < 21$ AMeV in order to extract the 
coefficient 
$\beta$ as a function of the temperature from the damping 
of the collective 
quadrupole oscillations of the formed di-nuclear system (DNS).
Density dependent effective interactions of Skyrme type 
($SKM$ \cite{kri90,bar98}) and an averaged free nucleon-nucleon 
cross section, $\sigma=40mb$ (\cite{bar98}) have been
used.

The structure of the work is as follows. In Sect. 2 the 
procedure
of the extraction of the reduced damping coefficient 
$\beta$ from
the time dependence of the quadrupole moment given by the 
BNV model is 
described. In Sect. 3 we present a novel method for 
determination
of the temperature based on the energy conservation and 
on 
the BNV evolution of the potential energy. Sect. 4 
contains 
our results on the $\beta(T)$ and their interpretation in 
terms
of two-body and one-body dissipation mechanisms.  
Summary and
conclusions are given in Sect. 5.

\section{ Extraction of the reduced friction coefficient 
          from nuclear kinetic equations}

According to Eq. (\ref{beta}), the reduced friction 
coefficient $\beta$
can be, in principle, calculated directly, once the 
phase-space distribution
function $f({\bf r},{\bf p}, t)$ is known from the BNV 
output, since:
\begin{eqnarray}
E_{kin}^{coll}(t) &=& \int d {\bf r}\, 
               { m {\langle v({\bf r},t) \rangle}^2 \over 2 }\,
               \rho({\bf r},t)~,                
                                           \label{Ekincoll} \\
\langle {\bf v}({\bf r},t) \rangle &=& { 1 \over \rho }
\int d {\bf p}\, f({\bf r},{\bf p},t) { {\bf p} \over m }~,                    
                                           \label{vcoll} \\
\rho({\bf r},t) &=& \int d {\bf p}\, f({\bf r},{\bf p},t)~,
                                           \label{rho} \\
E_{pot}(t) &=& E_{pot}^{int}(t) + 
{3 \over 5} \int d {\bf r}\, \epsilon_F(\rho)\rho~, 
                                           \label{Epotcoll} \\
E_{pot}^{int}(t) &=& \int d {\bf r}\, \epsilon_{m.f.}(\rho)
+ E_{coul}(t)~,                                                   
                                           \label{Epotcollint} 
\end{eqnarray}
where $\epsilon_F(\rho) = 
\hbar^2/(2m)(3\pi^2\rho/2)^{2/3}$ is the Fermi
energy, $\epsilon_{m.f.}(\rho)$ is the nuclear mean field 
interaction
energy density and $E_{coul}$ is the Coulomb energy. 
From Eq.(\ref{Ekincoll}) we see that the beam energy is not
giving contribution to the collective kinetic energy.

However, in practice,
the calculation of the collective kinetic energy 
$E_{kin}$ is quite
ambiguous in the test particle technique due to the 
strong dependence on
the width of the gaussians representing the test 
particles \cite{BNV}.

We calculate then the coefficient $\beta$ from the time 
evolution of
the quadrupole moment of the DNS:
\begin{equation}
Q_{zz}(t) = \int d {\bf r}\, q_{zz}({\bf r}) \rho({\bf 
r},t),~~~~~
q_{zz}({\bf r}) = 2z^2 - x^2 - y^2~.                 
\label{qzz}
\end{equation}
For a damped periodical motion (see Appendix)
\begin{equation}
Q_{zz}(t) \propto \exp( -i \omega t ),~~~
\omega = \omega_R + i\omega_I,~~~
\omega_I < 0~,                                     
\label{osc}
\end{equation}
the coefficient $\beta$ is proportional to the imaginary 
part of the
frequency $\omega$:
\begin{equation}
                  \beta = -4\omega_I~.              
\label{beta1}
\end{equation}

The curves in Fig. 1 show the time evolution of $Q_{zz}$ 
in the case of central 
collisions of $^{64}$Ni + $^{238}$U at beam energies 
$E_{lab} = 
6.53 \div 20.53$ AMeV. The quadrupole moment quickly 
approaches
a minimum at $t = 100\div200$ fm/c. Afterwards the 
$Q_{zz}$ starts to grow 
again, and after a time interval $\Delta t = 50\div100$ 
fm/c it saturates.
This saturation of the $Q_{zz}$ is caused by an almost 
complete lost
of the collective energy of the DNS. One can, therefore,
qualitatively estimate the $\beta$ coefficient just 
assuming that 
$|\omega_I| \simeq \omega_R = 2\pi/t_{osc}$, where 
$t_{osc} \simeq 100$ fm/c is the
period of oscillations. That gives $\beta = 4 |\omega_I| 
\simeq 70 \cdot 
10^{21}$ s. This value is in agreement with the results 
obtained in 
Refs. \cite{Wilc96,Don99}. However, we would like to 
stress that in
Refs. \cite{Wilc96,Don99} the damping coefficient was 
extracted
from the outgoing part of the trajectory of the DNS from  
compact mononucleus shape to scission during a time 
interval 
$\sim 30000$ fm/c. 
Our analysis is concentrated on the part of the 
trajectory  
{\it in vicinity of the mononucleus shape} and the 
corresponding time scale
$\sim 300$ fm/c is much shorter.
  
In order to get the values of the coefficient $\beta$
we fit a part of the curve $Q_{zz}(t)$ to the function
\begin{equation}
Q_{zz}^{fit}(t) = {\cal A} + {\cal B} \sin( \omega_R t + 
\phi_0 ) 
                      \exp( -|\omega_I|t )~.              
 \label{qzzfit}
\end{equation}
The upper time limit of the fitting region is chosen at 
the second minimum
of the quadrupole moment. The lower time limit is given 
by the earliest 
time when the same value of the $Q_{zz}$ as in the second 
minimum is 
reached. This definition of the time limits, from one 
hand, 
corresponds approximately  to the selection of a full 
oscillation in 
vicinity of the compact mononucleus shape. On the other 
hand, the stage
of strong dissipation from the first minimum to the first 
maximum is completely 
in the fitting region.
The best fit functions of Eq. (\ref{qzzfit}) are 
shown 
by full dots in Fig.1.
The fitting parameters ${\cal A},~{\cal 
B},~\omega_R,~\phi_0,~\omega_I$ and 
the corresponding coefficient $\beta = 4|\omega_I|$ for 
central collisions 
of $^{64}$Ni + $^{238}$U at various beam energies are 
collected in the {\it Table}.
We see that the damping increases with the increasing 
collision energy until $E_{lab}=10.53$ AMeV.
This fact can be understood already from Fig. 1, since 
with increasing beam energy 
a larger 
$\Delta Q_{zz} = Q_{zz}^{saturation} - Q_{zz}^{minimum}$ 
is damped during
a shorter time interval. But at higher 
energies $E_{lab} > 10.53$ AMeV the damping starts to 
decrease, as we can see from the presence of
quadrupole vibrations at later times 
$t > 200$ fm/c
(Fig. 1). Eventually at $E_{lab} > 14.53$ AMeV a kind of saturation 
seems to be reached.

\section{ Determination of the temperature from BNV simulations}

A direct way to extract the temperature is to fit the 
local 
momentum distribution given by the BNV model to a $T \neq 0$
Fermi distribution.
However, in the case of low-energy nuclear collsions 
studied in present
work, this direct method is not appropriate just because 
its accuracy
of $\sim 1$ MeV is not enough. Therefore we will follow a
procedure based 
on the conservation of the energy and on the time 
evolution of the potential
energy given by the BNV. The thermal excitation energy of 
the DNS is 
(c.f. \cite{Molec}):
\begin{equation}
E_{therm}^* = 
E_{kin}^{c.m.} - E_{coul} - E_{rot} - \Delta E_{pot}~,    
  \label{Eexc}
\end{equation}
where $E_{kin}^{c.m.} = E_{lab} A_1 A_2 / (A_1+A_2)$ is 
the center-of-mass
kinetic energy, $E_{coul} = e^2 Z_1 Z_2 / (R_1 + R_2 + 
3.5)$ is the Coulomb 
energy, $R_i = 1.2 A_i^{1/3}$ (fm) $i=1,~2$ are the 
nuclear radii, 
$E_{rot} = \hbar^2 L^2 / (2\Theta)$ is the rotational 
energy with $L$
being the angular momentum (in $\hbar$ units) and 
$ \Theta = [ {2 \over 5}( A_1 R_1^2 + A_2 R_2^2 ) +
             A_1 A_2 / (A_1+A_2) (R_1+R_2)^2 ] m_{nuc} $ 
being the momentum
of inertia ($m_{nuc}$ is the nucleon mass). The last term
$\Delta E_{pot}$ in the r.h.s. of Eq.(\ref{Eexc}) is the 
difference 
between the values of the potential energy just before 
and after the overlapping 
of the density profiles.
The origin of this term is mostly from the sharp decrease 
of the nuclear
surface energy when the two nuclei touch each other. In 
Fig. 2 we show the
time dependence of the potential energy per nucleon in 
the central collision
of $^{64}$Ni + $^{238}$U at 10.53 AMeV. After some small 
increase until
$t \simeq 20$ fm/c due to the Coulomb contribution the 
total potential
energy quickly drops by about 1 MeV/nucleon reaching the 
minimum at
$t \simeq 60$ fm/c. We define $\Delta E_{pot}$ as
\begin{equation}
      \Delta E_{pot} = E_{pot}^{min} - E_{pot}^{max}~,  
                                           \label{depot}
\end{equation}
where $E_{pot}^{min}$ and $E_{pot}^{max}$ are the first 
minimum and
the first maximum of the potential energy. 

In the {\it Table} we report the values of 
the potential
energy "jumps" $\Delta E_{pot}$, the excitation energies 
and corresponding
temperatures 
$T=\sqrt{E_{therm}^*/a},~a=A/8~\mbox{MeV}^{-1}$ 
for the $^{64}$Ni + $^{238}$U central collisions at 
various beam 
energies\footnote{ In the case of the central collisions 
$E_{rot}=0$ 
in Eq. (\ref{Eexc}) }. Our 
temperatures are higher than those 
obtained in Ref. \cite{Wilc96}. In particular, for the 
Ni+U collision at 6.53 AMeV
we have $T=3.1$ MeV and the authors of Ref. \cite{Wilc96} 
report 
$T=2.4$ MeV. This difference is mainly explained by the fact, 
that the temperatures 
in Ref.\cite{Wilc96} are obtained from the outgoing stage 
of the reaction on much 
larger time scales as  compared to our study. A further
contribute to some temperature overshooting is coming from
pre-equilibrium emissions, neglected in the
energy balance Eq.(\ref{Eexc}). This effect, although expected
to increase with beam energy, can be still considered quite
small in this energy range.

\section{ Temperature dependence of the dissipation }

Fig. 3 shows the temperature dependence of the reduced 
friction 
coefficient $\beta$. The BNV results are shown by solid 
line with dots.
The coefficient $\beta$ first increases with temperature 
reaching the
maximum at $T \simeq 4.5$ MeV and then at higher 
temperatures it decreases. This signal is indeed quite
robust and cannot be related to some overestimation of
the temperatures at higher beam energies as discussed at
the end of the previous section.

One can interpret the results of the full transport calculations 
within the 
approach of Ref. \cite{KPS96}, based on the 
analytical solution
of the linearized Landau-Vlasov equation
\begin{equation}
\left( {\partial \over \partial t}
    + {{\bf p} \over m^*} {\partial \over \partial {\bf 
r}} \right)
\delta f({\bf r},{\bf p},t)
    - {\partial \delta U \over \partial {\bf r}}
{\partial f_{\rm eq}({\bf p}) \over \partial {\bf p}} = 
I_{coll}[\delta f]~,                       
                                            \label{lveq}
\end{equation}
where $\delta f$ and $f_0$ are the perturbation and the 
equilibrium 
value of the phase space distribution function, 
\begin{equation}
\delta U({\bf r,p},t) = {1 \over N(0)}
\int\,{g d{\bf p'} \over (2\pi\hbar)^3}\,
( F_0 + F_1 \hat p \hat p' ) \delta f({\bf r,p'},t)   
\label{delU}
\end{equation}
is the mean field perturbation ($\hat p \equiv {\bf 
p}/p$,
$\hat p' \equiv {\bf p'}/p'$), $g=4$ is the spin-isospin 
degeneracy
of a nucleon, $N(0)=gm^*p_F/(2\pi^2\hbar^3)$ is the 
level density
at zero temperature, $F_0$ and $F_1$ are the Landau 
parameters and
$m^* = m / (1 + F_1/3)$ is the effective mass. 

The collision integral in the r.h.s. of Eq. (\ref{lveq}) 
is taken in
the relaxation time approximation:
\begin{equation}
I_{coll}[\delta f] \simeq -{1 \over \tau} \delta f_{|l 
\geq 2}~,              
                                       \label{collint}
\end{equation}
where 
\begin{equation}
\delta f_{|l \geq 2}({\bf p}) \equiv
\sum_{l \geq 2} \sum_{m=-l}^l\, 
Y_{lm}(\hat p) \int\, d\Omega_{\hat p'}\, Y^*_{lm}(\hat 
p')
\, \delta f({\bf p'})_{|p'=p}                   
                                       \label{delflge2}
\end{equation}
is the part of the perturbation containing the quadrupole 
and higher 
multipolarity distortions of the Fermi surface. The 
effective relaxation
time $\tau$ includes two- and one-body dissipation 
contributions:
\begin{equation}
\tau^{-1} = \tau^{-1}_{2body} + \tau^{-1}_{1body}~.      
\label{tau}
\end{equation}
The relaxation time $\tau_{2body}$ was calculated in Ref. 
\cite{LCBD99}
for various choices of a nucleon-nucleon scattering cross 
section:
\begin{equation}
\tau^{-1}_{2body} = T^2/\kappa                           
\label{tau2}
\end{equation}
with $\kappa \simeq 1900$ MeV$^2$fm/c for the isotropic 
energy independent
isospin-averaged nucleon-nucleon scattering cross section 
$\sigma_{NN}=40$ mb. For the one-body relaxation time we 
have used the
wall-and-window formula (c.f. Ref. \cite{KPS96}):
\begin{equation}
\tau^{-1}_{1body} = { \overline{v} \over 2 R \xi }~,     
\label{tau1}
\end{equation}
where $R = 1.2 A^{1/3}$ ($A = A_1 + A_2$) is the radius 
of a mononucleus
composed of the two colliding nuclei, 
\begin{equation}
\overline{v} = {3 \over 4} v_F 
\left[
        1 + {\pi^2 \over 6} \left( {T \over \epsilon_F} 
\right)^2
\right]                                                  
\label{vaver}
\end{equation}
is an average velocity of nucleons, and $\xi$ is a 
numerical
factor which depends on the multipolarity and on the 
isospin of
a collective mode. We have chosen a value $\xi=1.85$, 
which corresponds
to the isoscalar quadrupole mode in the scaling wall 
model
(see Ref. \cite{KPS96} and refs. therein). 

The solution of Eq. (\ref{lveq}) inside a nucleus with 
uniform 
nonperturbed density can be found as a superposition of 
plane waves.
In this case, as it was shown in Ref. \cite{KPS95}, the 
intrinsic
width of a giant multipole resonance
$\Gamma = 2\hbar/\tau_{rel}$, where $\tau_{rel}$ is 
the relaxation
time of the distribution function $\delta f$ ($\delta f 
\propto 
\exp( -i \omega t )$, $\omega = \omega_R + i\omega_I$, 
$\omega_I = -1/\tau_{rel}$),  can be
expressed as
\begin{equation}
\Gamma \simeq 2 q \hbar \omega_R \, 
{ \omega_R \tau \over 1 + q (\omega_R\tau)^2 }            
\label{gammaint}
\end{equation}
with $q = [ {5 \over 2} (1+F_0) (1+F_1/3) ]^{-1}$.  
Eq. (\ref{gammaint}) describes both well known \cite{AK} 
zero sound
($\Gamma \propto \tau^{-1}$, $\omega_R\tau \gg 1$) and 
first sound
($\Gamma \propto \tau$, $\omega_R\tau \ll 1$) regimes.
In our calculations we have put the Landau parameters 
$F_0=0.2$,
$F_1=0$. The frequency has been chosen as $\hbar\omega_R = 
64.7 A^{-1/3}$
MeV ($A = A_1 + A_2$) corresponding to the giant 
quadrupole resonance.

The solid line in Fig. 3 shows the friction coefficient 
\begin{equation}
      \beta_{ZFST} = 2 \Gamma / \hbar                   
\label{betaZFST}
\end{equation}
with $\Gamma$ given by Eq. (\ref{gammaint}). We see that 
Eqs. (\ref{gammaint}), (\ref{betaZFST}) agree qualitatively
with the BNV model, but the analytical calculation gives
a more smeared-out transition between the two sounds. 

In the limit of the zero sound the coefficient $\beta$ is
\begin{equation}
      \beta_{ZS} = 4/\tau = 
      \beta_{ZS}^{2body} + \beta_{ZS}^{1body}~,          
\label{betaZS}
\end{equation}
where $\beta_{ZS}^{2body} = 4/\tau_{2body}$,
      $\beta_{ZS}^{1body} = 4/\tau_{1body}$.
This simple formula (dot-dashed line in Fig. 3) is quite 
close to the BNV results at $T \leq 4.5$ MeV. However, either 
the two-body (dashed line) or one-body (dotted line) contributions 
taken separately are strongly underpredicting 
the BNV friction coefficient.

In order to better understand the relative importance of 
the two-body and one-body mechanisms we have also performed the BNV 
calculations at $E_{lab} = 6.53,~8.53,~10.53,~12.53$ and $14.53$ AMeV
switching off the collision term, corresponding to a 
pure Vlasov evolution. Fig. 4 shows the comparison of the $Q_{zz}$ 
time evolution with and without collision term for the Ni+U collision at 
14.53 AMeV. In the case without collisions the damping is much reduced: 
we observe several large-amplitude oscillations of the quadrupole 
moment. We have also checked that in the Vlasov evolution the oscillations 
at later times, $t > 200$ fm/c, are  present 
for all the other studied beam energies. 
In Fig. 3, the coefficient $\beta$  in the case of 
collisionless dynamics is shown by full squares. The damping slowly 
increases with temperature and saturates at $T = 5$ MeV. 
As expected, the system is always in the region of the zero sound. 
The absolute value of the reduced friction coefficient in the 
collisionless case is $3\div5$ times less than for a full BNV 
calculation and close to the wall-and-window value $4/\tau_{1body}$.

\section{ Summary and conclusions }

The reduced friction coefficient was extracted for the 
initial stage
($\sim 300$ fm/c) of central $^{64}$Ni + $^{238}$U 
collisions at
beam energies $E_{lab}=6\div21$ AMeV in the framework of 
the
BNV transport model. The quadrupole moment $Q_{zz}$ has been chosen 
as a 
relevant collective variable. The quadrupole time 
evolution
shows overdamped oscillations with a damping 
rate 
proportional to the friction coefficient. Two-body 
collisions play
a major role in the damping of $Q_{zz}$.

As a function of the beam energy, the damping rate has a 
maximum at 
$E_{lab} \simeq 10$ AMeV. The corresponding temperature 
of the DNS, 
neglecting particle emissions, is 4.6 MeV. We 
interpret, therefore, 
this temperature as a transition temperature from the 
zero-to-first sound 
propagation.

This result agrees with earlier calculations
of V.M. Kolomietz et al. (Ref. \cite{KPS96}), where a 
value of the
transition temperature $4\div5$ MeV was deduced on the basis of 
the analytical
solution of the linearized BUU equation in the case of 
isoscalar giant 
resonances of multipolarities $l=0$ and 2 in a hot 
nucleus.

Our calculations are overpredicting the transition 
temperature 
$2\div2.5$ MeV that follows from the analysis of the 
prescission neutron
multiplicities by J. Wilczy$\rm\acute n$ski et al. (Ref. 
\cite{Wilc96}). We have to stress that the considered collective modes
are different, in our analysis quadrupole oscillations in the entrance
channel dynamics while in the ref.\cite{Wilc96} the fission mode in the
exit channel. In particular a relevant variance is on the time scales of 
the two modes, 
with a much larger proper time for the fission dynamics.
Following the simple condition $\omega_R\tau \simeq 1$ for
the transition, we roughly get a $T_{tr} \propto \sqrt{\omega_R}$
and then we can expect quite smaller transition temperatures for the
fission modes. Moreover, as already stressed before, likely our
temperature assignements are a little overestimated, particularly for
the higher beam energies, due to the lack of pre-equilibrium
emissions in the energy balance.

Looking at the BNV simulations, the transition from zero-to-first sound 
appears as a presence 
of quadrupole vibrations at relatively late times, $t > 200$ 
fm/c 
(see Fig. 1), for beam energies above 10 AMeV. These 
vibrations would
accelerate the fission of the produced DNS that should be 
observed
experimentally, as an increase of fast-fission cross sections.

\section*{ Acknowledgements }

Authors are grateful to Dr. G.G. Adamian 
for many fruitful discussions and useful advices. Stimulating 
discussions with Prof. D.M. Brink are gratefully acknowledged.

\section*{ Appendix }

In this Appendix we derive the relation (\ref{beta1}) between the 
imaginary part
$\omega_I$ of the complex frequency $\omega = \omega_R + 
i\omega_I$ and
the reduced damping coefficient $\beta$ of Eq. 
(\ref{beta}).

From the Eq. (\ref{qzz}) and the continuity equation 
$\dot{\rho} = -{\bf\nabla}({\bf v}\rho)$ we get:
\begin{equation}
      \dot{Q}_{zz} = 
      \int \rho {\bf v \nabla} q_{zz} d^3 r~.           
\label{qzztd}
\end{equation}
In the case of the small-amplitude damped periodical 
motion of the
kind
\begin{equation}
      {\bf v}({\bf r},t) = 
      \delta {\bf v}_0({\bf r}) \exp( -i \omega t )     
\label{vrt}
\end{equation}
the Eq. (\ref{qzztd}) can be linearized with respect to 
small values of
$\delta v_0$:
\begin{equation}
      \dot{Q}_{zz} \simeq 
      \left[ 
      \rho_0 \int \delta {\bf v}_0({\bf r}){\bf\nabla} 
q_{zz} d^3 r
      \right] \exp( -i \omega t ) \equiv 
      \dot{Q}_{zz}(t=0) \exp( -i \omega t )~,           
\label{qzztd1}
\end{equation}
where $\rho_0$ is the nonperturbed density ($\rho = 
\rho_0 + \delta\rho$).
The general solution of Eq. (\ref{qzztd1}) is
\begin{equation}
      Q_{zz}(t) = \mbox{const} + 
                  { i\dot{Q}_{zz}(0) \over \omega }
                  \exp( -i \omega t )~.                 
\label{qzz1}
\end{equation}
Therefore in a linear approximation, over $\delta v_0$,
the quadrupole
moment reveals oscillations with the same frequency 
$\omega = \omega_R + i\omega_I$ as the velocity field ${\bf 
v}({\bf r},t)$.

The time derivative of the collective kinetic energy 
(\ref{Ekincoll}) is:
\begin{eqnarray}
\dot{E}_{kin} & \simeq & 
m \rho_0 \int {\rm Re}({\bf v}) {\rm Re}({\bf \dot{v}}) 
d^3 r = \nonumber \\
& & m \rho_0 \left[ \int \delta v_0^2({\bf r}) d^3 r 
\right]
\cos(\omega_Rt) ( \omega_I\cos(\omega_Rt) - 
\omega_R\sin(\omega_Rt) )
\exp( 2\omega_It )~.                                
\label{Ekintd}
\end{eqnarray}
Averaging Eq. (\ref{Ekintd}) over the period of 
oscillations we come
to the relation:
\begin{equation}
\overline{ \dot{E}_{kin} } \simeq 
\omega_I \rho_0 m \int \overline{ ({\rm Re}({\bf v}))^2 } 
d^3 r =
2 \omega_I \overline{ E_{kin} }~,                    
\label{Ekintdav}
\end{equation}
where we have dropped the therm 
$\propto \cos(\omega_Rt) \sin(\omega_Rt) \exp( 2\omega_It 
)$ 
which changes the sign during the period of oscillation. 
According to the virial theorem for harmonic oscillators
\begin{equation}
\overline{ E_{kin} } = \overline{ E_{pot} } = 
{ 1 \over 2 } \overline{E}~,                        
\label{virtheor}
\end{equation}
where $E$ is the total collective energy. Therefore
\begin{equation}
\overline{ \dot{E} } \simeq 4 \omega_I \overline{ E_{kin} 
}~.          
\label{Etottd}
\end{equation}

\newpage

\begin{description}
\item[Table] Fit parameters from Eq. (\ref{qzzfit}); 
potential energy
"jumps" $\Delta E_{pot}$; excitation energies and 
temperatures of 
the DNS Eqs. (\ref{Eexc}),(\ref{depot}); for central 
collisions
$^{64}$Ni + $^{238}$U at various beam energies $E_{lab}$.
\end{description}

\vspace{0.5cm}

\begin{center}
\begin{tabular}{|c|c|c|c|c|c|c|c|c|}
\hline
$E_{lab}$  &  ${\cal A}$  &  ${\cal B}$  &  $\omega_R$  & 
 $\phi_0$  &  
$\omega_I$ &  -$\Delta E_{pot}$  &  $E^*$  &  T  \\
(AMeV)     &  (Afm$^2$)   &  (Afm$^2$)   &  c/fm        & 
           &
c/fm       &  AMeV              &  AMeV   &  MeV \\
\hline
6.53       &  33.81       & 24.34        & 0.024        & 
-6.537     &
0.0133     &  0.88        &  1.2         &  3.1 \\
\hline
8.53       &  25.60       &  39.94       & 0.030        & 
-6.169     &
0.0175     &  1.28        &  1.9         &  3.9 \\
\hline
10.53      &  21.46       &  38.45       & 0.036        & 
-6.197     &
0.0180     &  1.68        &  2.7         &  4.6 \\
\hline
12.53      &  18.34       &  31.13       & 0.038        & 
-6.091     &
0.0163     &  2.10        &  3.4         &  5.2 \\
\hline
14.53      &  16.42       &  27.03       & 0.039        & 
-5.906     &
0.0147     &  2.28        &  3.9         &  5.6 \\
\hline
16.53      &  14.33       &  23.50       & 0.040        & 
-5.843     &
0.0137     &  2.46        &  4.4         &  6.0 \\
\hline
18.53      &  12.89       &  26.73       & 0.039        &
-5.589     &
0.0141     &  2.68        &  5.0         &  6.3 \\
\hline
20.53      &  11.14       &  28.05       & 0.038        & 
-5.486     &
0.0147     &  2.74        &  5.4         &  6.6 \\
\hline
\end{tabular}
\end{center}

\newpage

\section*{ Figure captions }

\begin{description}

\item[Fig. 1] Solid lines -- time dependence of the 
quadrupole
moment $Q_{zz}$ for $^{64}$Ni + $^{238}$U central 
collisions at 
beam energies (from top to bottom) $E_{lab}=$ 6.53, 
8.53,
10.53, 12.53, 14.53, 16.53 and 20.53 AMeV. The value of
the quadrupole moment at t=0 is always 85 fm$^2$/nucleon.
Full circles show the best fits obtained using Eq. 
(\ref{qzzfit}) for 
each collision energy. See text after  Eq. (\ref{qzzfit})
for a definition of the fit regions.

\item[Fig. 2] Potential interaction energy per nucleon 
Eq. (\ref{Epotcollint}) as a function of time 
for the $^{64}$Ni(10.53 AMeV) + $^{238}$U central 
collision. 

\item[Fig. 3] Temperature dependence of the reduced 
friction coefficient
$\beta$. BNV calculations with (without) collision term are
shown by 
full circles (squares) connected with thin solid line 
to guide eye. Errorbars are due to the ambiguity caused by a
finite number of test particles. 
The analytical result of Ref. \cite{KPS96} for the Giant 
Quadrupole
Resonance in a hot nucleus (see Eqs. (\ref{gammaint}), 
(\ref{betaZFST}))
is shown by thick solid line. The two- and one-body 
contributions and
their sum (see Eq. (\ref{betaZS})) in the zero sound 
limit are shown by
dashed, dotted and dash-dotted lines respectively.

\item[Fig. 4] Comparison of the time evolution of the 
quadrupole moment
for $^{64}$Ni(14.53 AMeV) + $^{238}$U central collision 
calculated with 
(solid line) and without (dashed line) collision term.

\end{description}

\clearpage
\thispagestyle{empty}

\begin{figure}[btp]
\psfig{figure=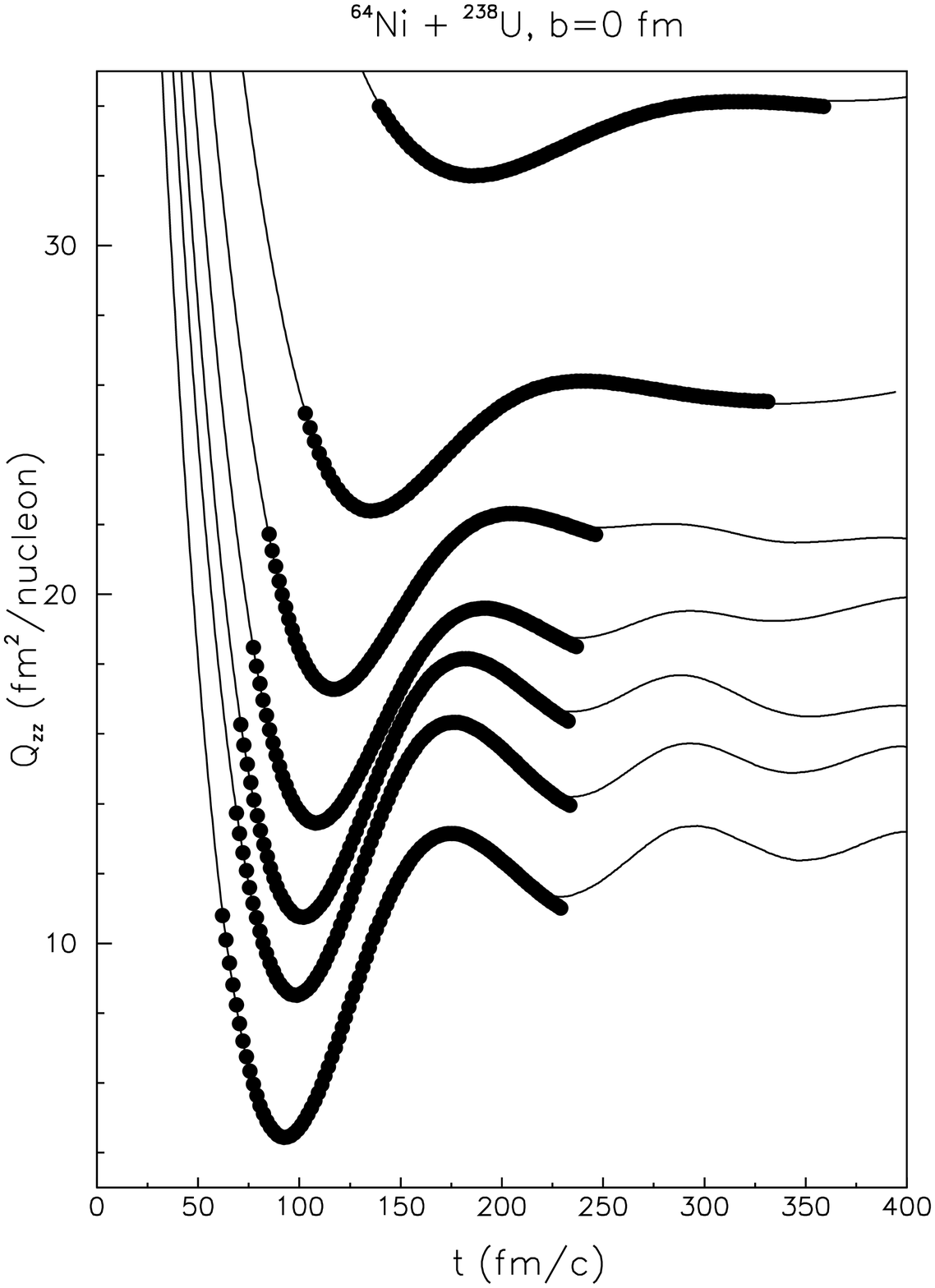,width=\textwidth}
\caption{ }
\end{figure}

\clearpage
\thispagestyle{empty}

\begin{figure}[btp]
\psfig{figure=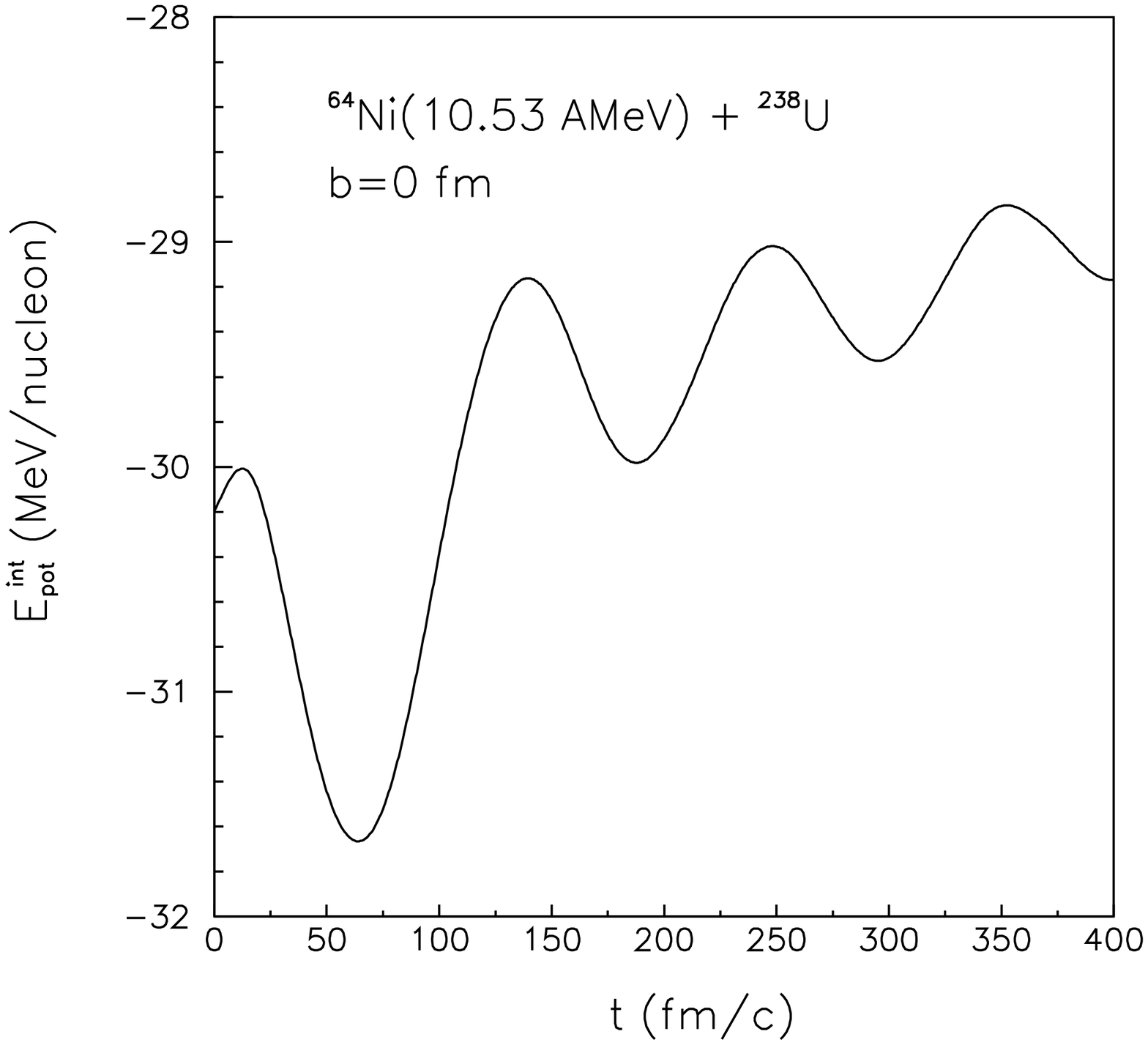,width=\textwidth}
\caption{ }
\end{figure}

\clearpage
\thispagestyle{empty}

\begin{figure}[btp]
\psfig{figure=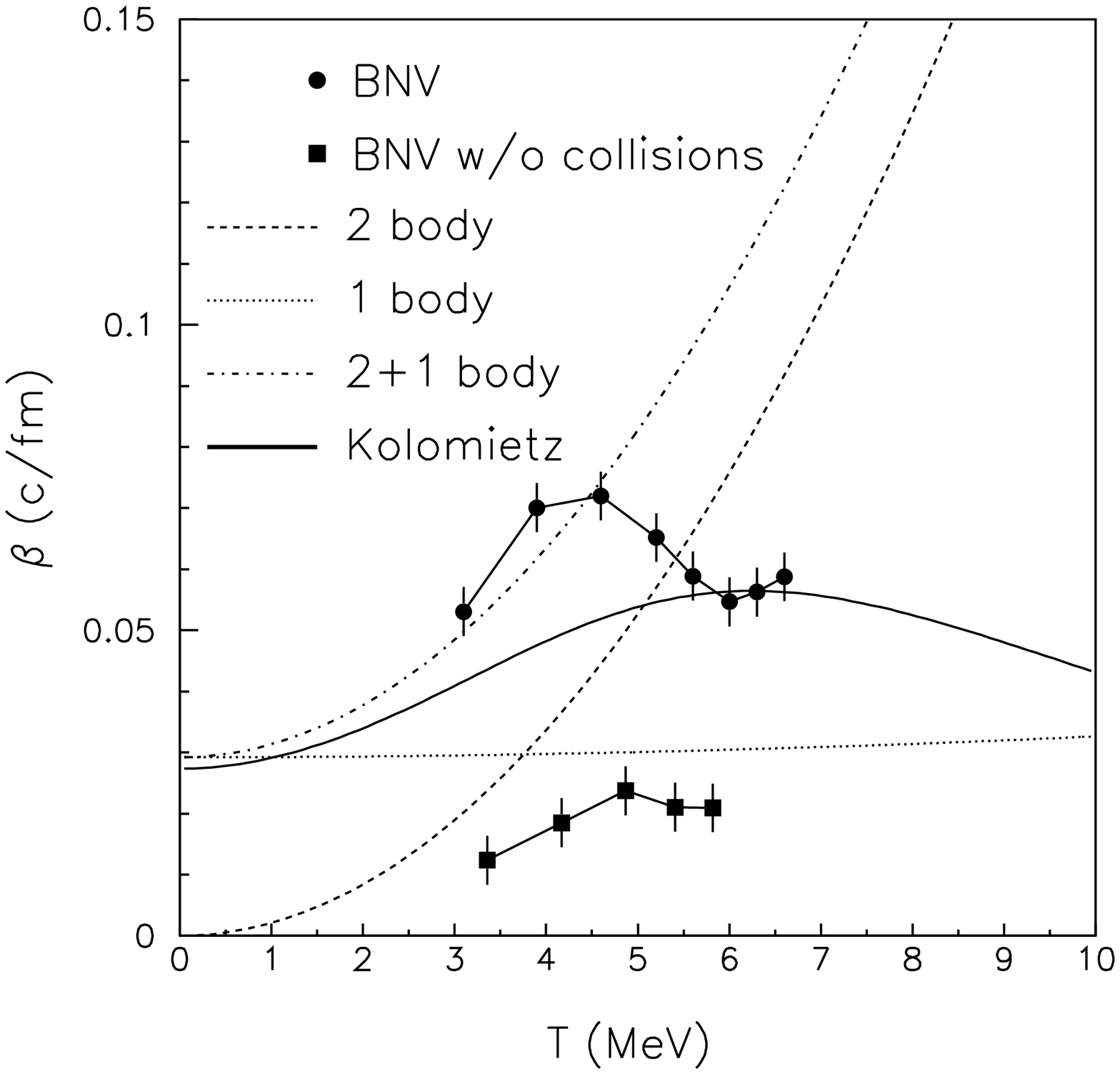,width=\textwidth}
\caption{ }
\end{figure}

\clearpage
\thispagestyle{empty}

\begin{figure}[btp]
\psfig{figure=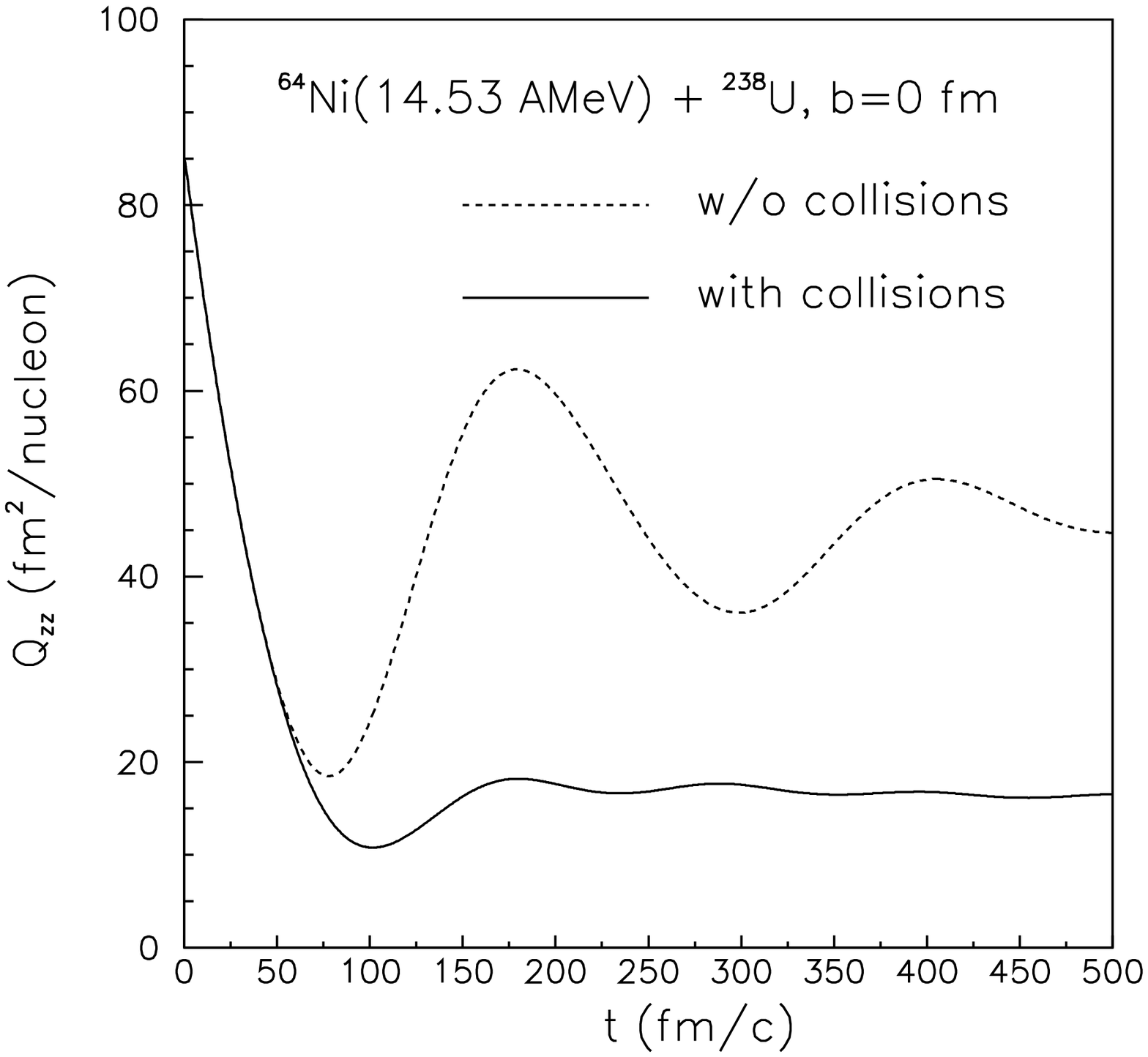,width=\textwidth}
\caption{ }
\end{figure}

\end{document}